\documentclass[journal=jacsat,manuscript=article]{achemso}

\usepackage[version=3]{mhchem} 
\usepackage{times}
\usepackage{bm}
\usepackage{amsmath,amssymb}
\author{Sathwik Bharadwaj}
\affiliation{Elmore Family School of Electrical and Computer Engineering, Purdue University, West Lafayette, Indiana 47907, USA.}
\affiliation{Worcester Polytechnic Institute, Worcester, Massachusetts, 01609, USA.}
\email{sathwik@wpi.edu}
\author{Makoto Schreiber}
\affiliation{Nanotechnology Research Centre, National Research Council of Canada, Edmonton, Alberta T6G 2M9,
Canada.}
\author{Jungho Mun}
\affiliation{Elmore Family School of Electrical and Computer Engineering, Purdue University, West Lafayette, Indiana 47907, USA.}
\author{Sam Ruttiman}
\affiliation{Nanotechnology Research Centre, National Research Council of Canada, Edmonton, Alberta T6G 2M9,
Canada.}
\author{Pronoy Das}
\affiliation{Elmore Family School of Electrical and Computer Engineering, Purdue University, West Lafayette, Indiana 47907, USA.}
\author{Misa Hayashida}
\affiliation{Nanotechnology Research Centre, National Research Council of Canada, Edmonton, Alberta T6G 2M9,
Canada.}
\author{Marek Malac}
\affiliation{Nanotechnology Research Centre, National Research Council of Canada, Edmonton, Alberta T6G 2M9,
Canada.}
\email{marek.malac@nrc-cnrc.gc.ca / mmalac@ualberta.ca}
\author{Peter Nordlander}
\affiliation{Department of Materials Science and NanoEngineering, Department of Electrical and Computer Engineering, and  Department of Physics and Astronomy, Rice University, Houston, TX, USA.}
\author{Zubin Jacob}
\affiliation{Elmore Family School of Electrical and Computer Engineering, Purdue University, West Lafayette, Indiana 47907, USA.}
\email{zjacob@purdue.edu}
\title[An \textsf{achemso} demo]{Observation of Crystalline Nonlocal Volume Plasmon Waves}
\begin{document} 
\begin{abstract}
In plasmonics, nonlocal effects arise when the material response to optical excitations is strongly dependent on the spatial correlations of the excitation. It is well known that a classical free electron gas system supports local Drude volume plasmon waves. Whereas a compressible quantum electron gas system sustains hydrodynamic volume plasmons with nonlocal dispersion isotropic across all high-symmetry directions. Here, distinct from Drude and Hydrodynamic plasmon waves, we present the first observation of crystalline nonlocal volume plasmon waves. We use transmission-based momentum-resolved electron energy loss spectroscopy to measure the volume plasmon dispersion of silicon along all the fundamental symmetry axes, up to high momentum values ($q \sim 0.7$ reciprocal lattice units). We show that crystalline nonlocal plasmon waves have a prominent anisotropic dispersion with higher curvature along the light-mass ($\Gamma K$ \& $\Gamma L$) axes, compared to the heavy-mass ($\Gamma X$) axis. We unveil the origin of this phenomenon by experimentally extracting the anisotropic Fermi velocities of silicon. Our work highlights an exquisite nonlocality-induced anisotropy of volume plasmon waves, providing pathways for probing many-body quantum effects at extreme momenta.
\end{abstract}
\baselineskip24pt
\maketitle
Plasmon waves were among the earliest known cases of collective excitation in solid-state systems \cite{Pines_plasmons}. When an equilibrium charge density of a material is perturbed, long-range Coulomb interactions of electrons generate self-sustaining plasmon waves. Plasmon waves in a material are classified into two categories: a) volume plasmon waves, which are the longitudinal oscillations of electrons that propagate through the bulk of a material, and b) surface plasmon waves, which are collective oscillations of free electrons at an interface of a material. The study of nonlocal effects and spatial dispersion of plasmon waves has gained significant recent interest in nanophotonics \cite{monticone2025roadmap}.  Nonlocal phenomena resulting from the quantum confinement of surface plasmon waves in noble metals have been extensively studied in the literature \cite{garcia2008nonlocal, luo2013surface, mortensen2014generalized, scholl2012quantum, poursoti2022deep}. However, volume plasmon waves are generally considered local and isotropic across all symmetry directions of a material.

\begin{figure}[th]
    \centering
    \includegraphics[width=\linewidth]{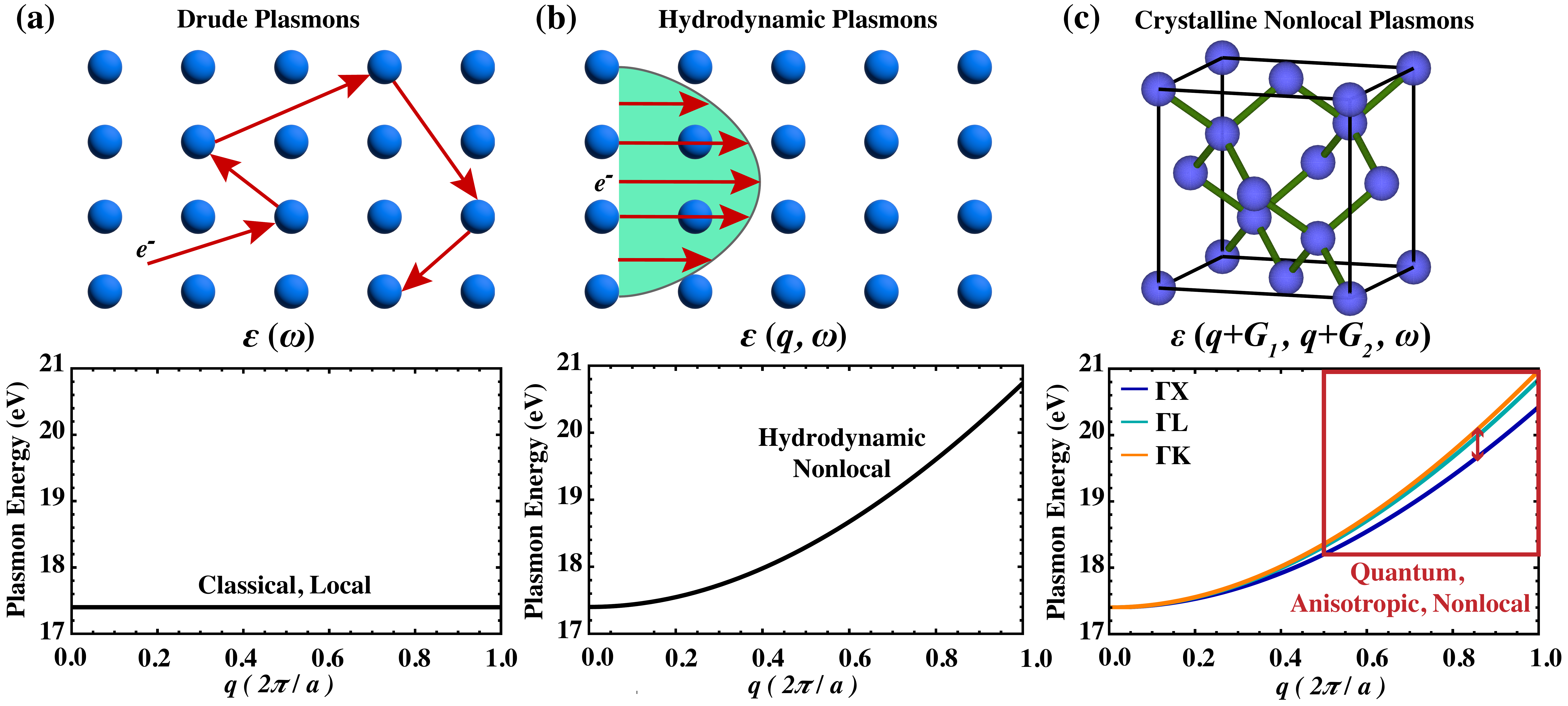}
    \caption{(a) Drude plasmons are supported by a system of free electron gas in a material, which can move randomly and collide with heavier nuclei. Drude plasmons display a zero-curvature dispersion with momentum $q$. (b) Hydrodynamic Plasmons are supported by a compressible electron gas system, obeying Pauli’s exclusion principle of electrons. Hydrodynamic plasmons exhibit an isotropic parabolic $q$-dependency across all symmetry axes. (c) Here, we introduce a class of crystalline nonlocal volume plasmon waves, which display both nonlocal and highly anisotropic dispersion. Curvature of the crystalline nonlocal plasmons strongly depends on the high-symmetry axes of a material.} 
    \label{fig:figure1}
\end{figure}

Classically, electrons in a material are considered a gas of free particles that can move randomly and collide with heavier ions. This system supports local Drude plasmon waves, with a zero curvature dispersion with momentum $q$ (Fig.~\ref{fig:figure1}(a)). Traditional electron energy loss spectroscopy (EELS) experiments can effectively capture the dispersion of Drude plasmons \cite{chen1975electron}. However, as fermions, electrons obey Pauli's exclusion principle, which makes the electron gas in a material a compressible system. A compressible electron gas system supports Hydrodynamic plasmon waves, which have a parabolic $q$-dependency across all symmetry directions of a material (Fig.~\ref{fig:figure1}(b)). With the development of EELS experiments in conjunction with transmission electron microscopy, the parabolic dispersion of hydrodynamic volume plasmon waves has been mapped in a variety of material systems, including silver \cite{zacharias1976dispersion}, aluminum \cite{shekhar2017momentum}, and Bi$_2$Se$_3$ \cite{PhysRevB.87.085126}. Here, distinct from conventional Drude and Hydrodynamic plasmon waves, we observe a class of crystalline nonlocal volume plasmon waves. 

We show that the dispersion curves of crystalline nonlocal plasmon waves are strongly anisotropic across the distinct high-symmetry axes of a material. We consider silicon (Si) as a prototype material and demonstrate both experimentally and theoretically that crystalline nonlocal plasmon waves have higher dispersion curvature along the light-mass ($\Gamma K$ \& $\Gamma L$) axes, compared to the heavy-mass ($\Gamma X$) axis (Fig.~\ref{fig:figure1}(c)). We unveil the origin of this phenomenon by extracting the anisotropic Fermi velocities of Si along all distinct symmetry axes. 

Nonlocal effects predominate in the dispersion of surface plasmon waves when the length scales of metallic nanostructures approach a few nanometers. Nanostructures, such as picocavities \cite{urbieta2018atomic}, coupled nanoparticles \cite{ciraci2012probing}, and the two-dimensional (2D) superlattice of nanoparticles \cite{shen2017optical} have been shown to exhibit significant nonlocal quantum effects \cite{teperik2013quantum, PhysRevLett.110.263901}. These mesoscopic phenomena occurring at the interface of metal and dielectric materials can be theoretically characterized by computing the Feibelman $d$-parameters \cite{christensen2017quantum}. Both EELS and cathodoluminescence measurements can be used to experimentally test the nonlocal effects of surface plasmon waves by retrieving Feibelman $d$-parameters \cite{gonccalves2023interrogating}. We note that, contrary to the crystalline nonlocal plasmon waves discussed here, the nonlocal effects in surface plasmon waves are imposed by subnanoscale geometry rather than the intrinsic spatial anisotropy of materials within the crystal lattice.   

Our $q$-EELS technique maps the energy-momentum dispersion of volume plasmon waves with high accuracy. Contrary to reflection-based EELS \cite{abbamonte2025collective} measurements, which can map optical excitations only up to $q \sim 0.1$ reciprocal lattice units, our transmission-based EELS allows us to measure nonlocal plasmonic phenomena at extremely large momenta ($q \sim 0.7$ reciprocal lattice units). Despite earlier attempts to map the dispersion of volume plasmon waves \cite{Raether}, the experimental precision at high momentum values has limited the observation of crystalline nonlocal plasmon waves. There has also been growing recent interest in understanding the dispersion of volume plasmon waves in dark matter research, particularly for low-mass dark matter detection \cite{PhysRevLett.127.151802}. Our observation of the anisotropic nature of crystalline nonlocal plasmon waves may provide pathways for directional detection of low-mass dark matter through daily modulation of the time-dependent event rate \cite{PhysRevD.108.015015}. 

\begin{figure}[th]
    \centering
    \includegraphics[width=\linewidth]{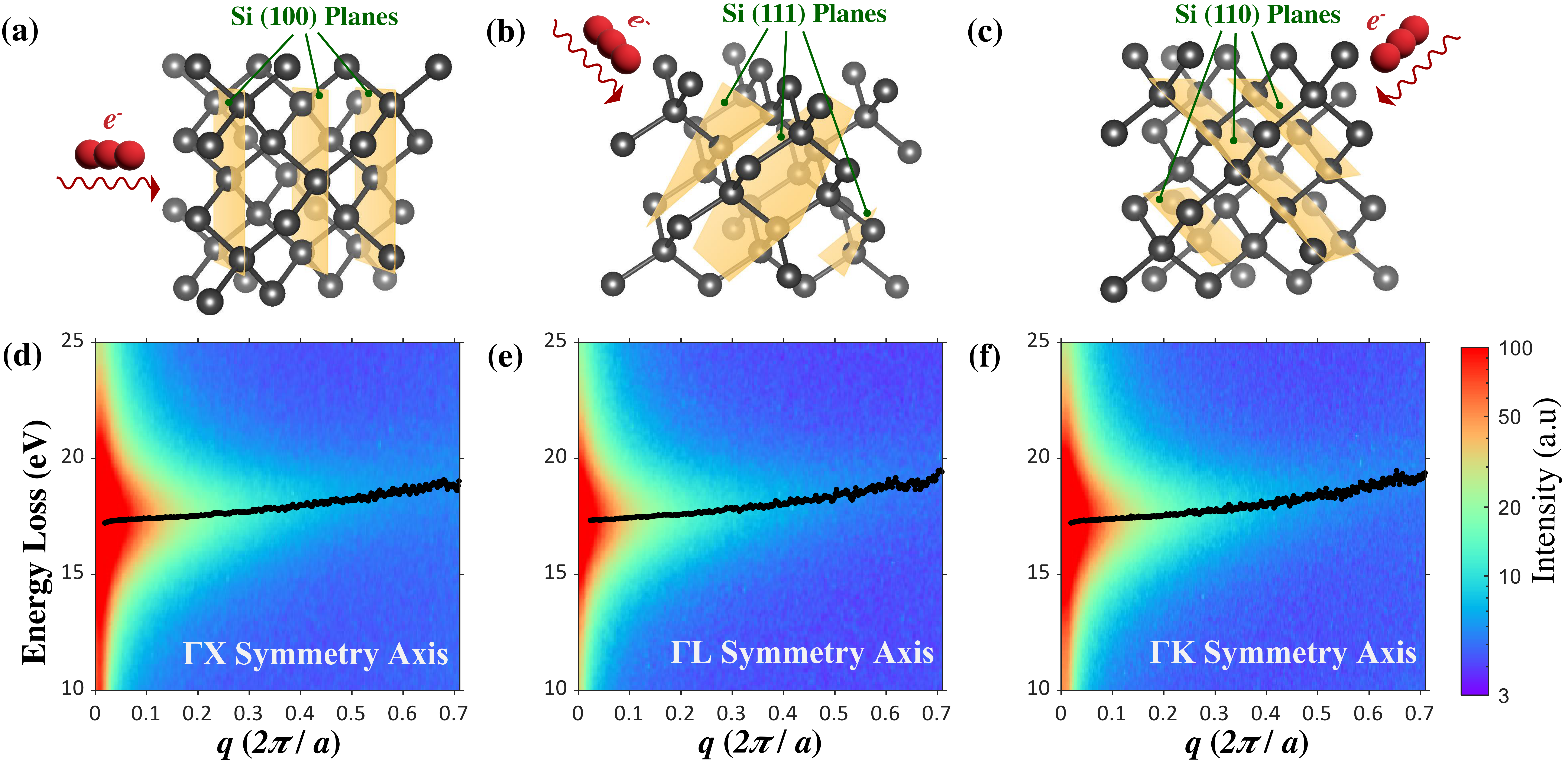}
    \caption{(a) Schematic illustration of a parallel electron-beam incident on the crystalline planes of silicon along (a) $\left[100\right]$ ($\Gamma X$), (b) $\left[111\right]$ ($\Gamma L$), and $\left[110\right]$ ($\Gamma K$) symmetry axes. (d, e, f) The momentum-resolved EELS signal for three different high symmetry directions in silicon. The signal intensity is visualized on a color axis on a logarithmic scale; the extracted maxima for the $q$-EELS signal correspond to the energy of crystalline nonlocal volume plasmon waves, represented by black dots.}
    \label{fig:figure2}
\end{figure}

In Fig.~\ref{fig:figure2}, we have plotted the $q$-EELS signal in Si for the three distinct high-symmetry directions. The extracted maxima for the EELS signal correspond to the crystalline nonlocal volume plasmon waves (shown as black dots), and the signal intensity is plotted as a color map on a logarithmic scale. The maxima of the EELS signal were obtained by fitting the raw signal spectra to the asymmetric pseudo-Voigt profile. The pseudo-Voigt spectral profile is given as a linear combination of Gaussian and Lorentzian profiles:
\begin{equation}
f(\omega; \omega_0, \gamma, \nu, \alpha) = \nu\,f_G(\omega; \omega_0, \gamma_\alpha) + (1-\nu)\,f_L(\omega; \omega_0, \gamma_\alpha),
\end{equation}
where,
\begin{equation}
f_G(\omega; \omega_0, \gamma_\alpha) = \frac{1}{\gamma_\alpha}\sqrt{\frac{4\ln{2}}{\pi}}\exp\left[{-4\left(\ln{2}\right)\frac{(\omega-\omega_0)^2}{\gamma^2_\alpha}}\right],
\end{equation}
and
\begin{equation}
f_L(\omega; \omega_0, \gamma_\alpha) = \frac{1}{\pi}\left[\frac{(\gamma_\alpha/2)}{(\omega-\omega_0)^2+(\gamma_\alpha/2)^2}\right].
\end{equation}
Here, $\omega_0$ is the center frequency, $\gamma$ is the full width half maximum, and $\nu$ is the ratio of the Gaussian and Lorentzian profiles. To incorporate asymmetric profiles, we considered the parameter \mbox{$\gamma_\alpha = 2\gamma/(1+\exp{\left[\alpha(\omega-\omega_0)\right]})$}, where $\alpha$ is the asymmetry parameter \cite{stancik2008simple}. 

To quantify the anisotropy of crystalline nonlocal plasmon waves, let us assume that along each high-symmetry axis, we have an isotropic three-dimensional electron gas (3DEG). We obtain a dispersion relation for volume plasmon waves within a long-wavelength approximation, given by
\begin{equation}
\omega^2(q) =  \omega_P^2 + \frac{3}{5} v_F^2\,q^2,
\end{equation}
where $v_F$ is the Fermi velocity and $\omega_P$ is the plasmon frequency at $q = 0$. The volume plasmon dispersion was obtained by fitting the maxima of the electron energy loss signal to the dispersion model. We fit the plasmon dispersion for three different high-symmetry directions, $\Gamma X$, $\Gamma L$, and $\Gamma K$, separately. In Table~\ref{tab:tab1}, we have shown the parameters of Fermi velocity ($v_F$) and Fermi energy ($E_F$) obtained by fitting the 3DEG model along the high-symmetry directions. We observe that the high-mass direction ($\Gamma X$) has a lower Fermi velocity compared to the low-mass directions $\Gamma K$ and $\Gamma L$. Our experimental results are also consistent with recent theoretical work, which considered the case of volume plasmon dispersion with explicit mass anisotropy in a three-dimensional electron gas system \cite{PhysRevB.103.L041303}.
\begin{table}[t!]
    \centering
    \begin{tabular}{|c|c|c|c|}
    \hline\hline
        Direction & $\hbar\omega_P\,$(eV) & $v_F\,$(km/s)  & $E_F\,$(eV) \\
        \hline\hline
        $\Gamma X$ & $17.4\pm0.01$ & $1576\pm14$  &   $9.31$\\
        $\Gamma K$ & $17.4\pm0.02$ & $1705\pm18$ &  $10.07$\\
        $\Gamma L$ & $17.4\pm0.02$ & $1707\pm16$ &  $10.08$\\
        \hline\hline
    \end{tabular}
    \caption{Fermi velocity ($v_F$) and Fermi energy ($E_F$) parameters obtained by fitting the $q$-EELS experimental data with the isotropic electron gas model along high symmetry directions. We observe that the high-mass direction ($\Gamma X$) has a lower Fermi velocity when compared to lower-mass directions $\Gamma K$ and $\Gamma L$.}
    \label{tab:tab1}
\end{table}

Further, we theoretically analyze the dispersion of volume plasmons within the crystalline nonlocal framework, including deep microscopic momentum transfer processes \cite{PhysRevApplied.18.044065}. For a given wavevector ${\bf q}$ and frequency $\omega$, within the linear response theory, the induced potential $\delta V_{\rm ind}$ in a material due to an external potential $\delta V_{\rm ext}$ can be expressed as
\begin{equation}
\delta V_{\rm ext}({\bf q}+\bf{G}_1, \omega) = \sum_{\bm{G}_1,\bm{G}_2} \varepsilon_L(\bm{q}+\bm{G}_1, \bm{q}+\bm{G}_2, \omega)\,\delta V_{\rm ind}({\bf q}+\bf{G}_2, \omega),
\end{equation}
where $\bm{G}_1$ and $\bm{G}_2$ represent the reciprocal lattice vectors and $\varepsilon_L(\bm{q}+\bm{G}_1, \bm{q}+\bm{G}_2, \omega)$ is the deep-microscopic longitudinal dielectric response tensor. We have implemented this crystalline nonlocal theoretical framework within our recently developed PicoMax software \cite{Bharadwaj_optical_polarization}. Starting from the electronic band structure, PicoMax computes $\varepsilon_L(\bm{q}+\bm{G}_1, \bm{q}+\bm{G}_2, \omega)$, including all the momentum-exchange processes within the crystal lattice.  We note that the majority of density functional theory-based software compute the dielectric function in the long-wavelength limit $(q\rightarrow 0)$, which is insufficient to quantify the dispersion of volume plasmon waves at high momenta. The crystalline nonlocal quantum framework, implemented within PicoMax, allows us to compute \mbox{$\varepsilon_L(\bm{q}+\bm{G}_1, \bm{q}+\bm{G}_2, \omega)$} for all momenta within the Brillouin zone of a material, including all the crystal symmetries. Theoretical and computational details of these calculations are discussed elsewhere \cite{Bharadwaj_optical_polarization}. 

The crystalline nonlocal plasmon waves are self-sustaining charge oscillations produced by a longitudinal electric field in the absence of external charge densities. Hence, the condition for self-sustained volume plasmon waves in a material is given by
\begin{equation}\label{eq:plasmondispersion}
{\rm det}\left[\varepsilon_L(\bm{q}+\bm{G}_1, \bm{q}+\bm{G}_2, \omega)\right] = 0. 
\end{equation}
We note that in the continuum limit, one can reduce the above equation to the standard form
\begin{equation}
\varepsilon(\bm{q}, \omega) = 0,     
\end{equation}
which neglects the crystalline symmetries and the deep microscopic momentum-exchange processes within the crystal lattice. 

\begin{figure}[th]
    \centering
    \includegraphics[width=0.8\linewidth]{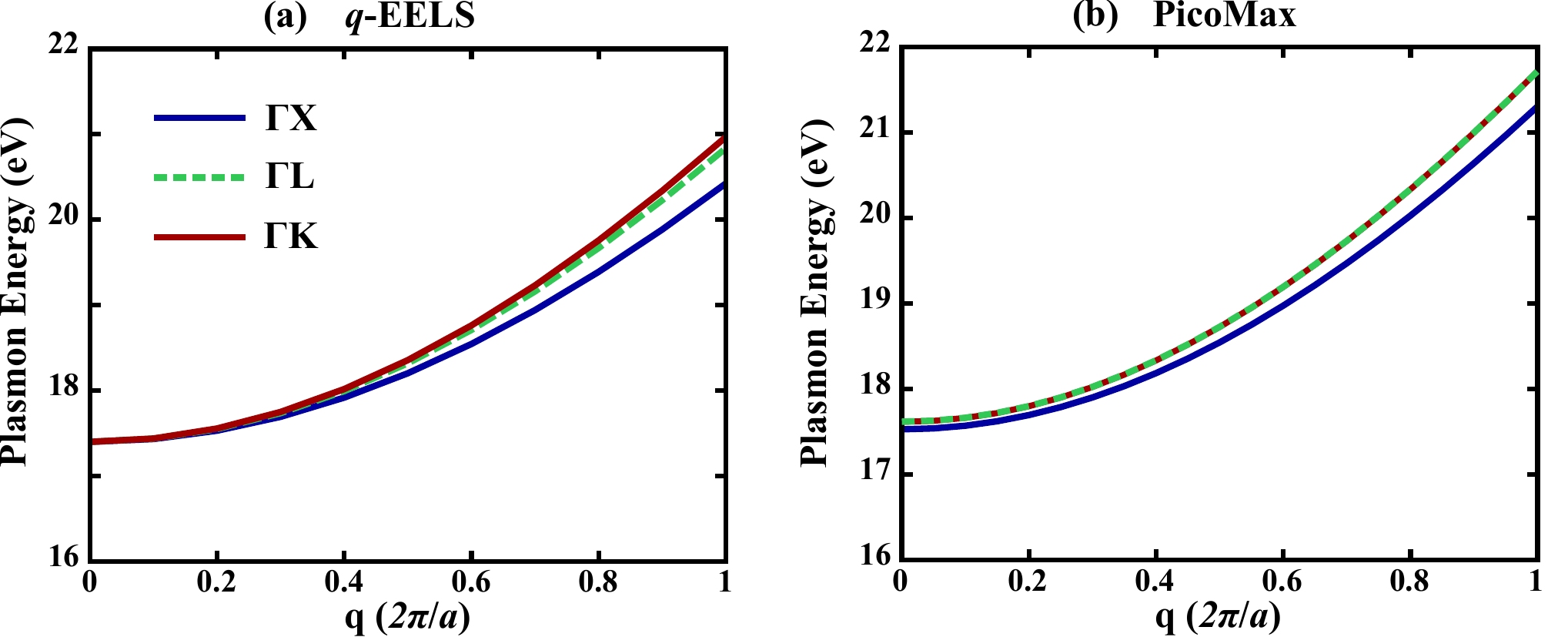}
    \caption{Comparison between theory and experiments for dispersion of crystalline nonlocal volume plasmon waves. (a) Experimental dispersion obtained by fitting the $q$-EELS data with the isotropic electron gas model. (b) Theoretical calculations were performed within the crystalline nonlocal framework implemented in PicoMax software along $\Gamma {\rm X}$, $\Gamma {\rm L}$, and $\Gamma {\rm K}$ high-symmetry directions.}
    \label{fig:figure3}
\end{figure}

By solving Eq.~(\ref{eq:plasmondispersion}), we obtain the dispersion curves for crystalline nonlocal plasmon waves. In Fig.~\ref{fig:figure3}, we observe a strong match between the $q$-EELS experimental measurements and PicoMax data in determining the anisotropy of crystalline nonlocal plasmon waves and the dependency of the dispersion curves on the high symmetry directions of silicon. 

In summary, we observed crystalline nonlocal plasmon waves in silicon using transmission-based momentum-resolved electron energy loss spectroscopy. We observed that crystalline nonlocal plasmon waves along the $\Gamma L$ and $\Gamma K$ axes exhibit a higher curvature compared to the $\Gamma X$ axis. The origin of this curvature difference is attributed to the higher Fermi velocity along the $\Gamma L$ and $\Gamma K$ axes compared to the $\Gamma X$ axis. We show that this experimental observation, through our $q$-EELS measurements, matches the crystalline nonlocal theoretical calculations implemented in our PicoMax software. Our study opens up opportunities for exploring bulk collective excitations of materials at extreme momenta.

\section*{Methods}
{\bf Momentum-resolved Electron Energy Loss Spectroscopy:} We prepared the single-crystal Si samples by focused ion beam (FIB) milling of single-crystal wafers. Two orientations of the Si samples were prepared with crystallographic orientation such that the incident electrons propagate along $[111]$ and $[100]$ directions. These two samples allowed us to obtain $q$-EELS spectra with momentum $q$ along the $\Gamma X$, $\Gamma K$, and $\Gamma L$ high-symmetry directions of the Brillouin zone. We selected the angular range of the $q$-EELS spectra such that Bragg reflections of 1$^{st}$ order are included in the $q$-EELS spectrum images. A rectangular angle-selecting slit was placed in front of the spectrometer to define a small region in the direction perpendicular to the angle-dispersion direction.

The intensity of the volume plasmon waves drops away from the zero-order beam. Hence, care must be taken to maintain an adequate signal-to-noise ratio (SNR) of the spectra while avoiding saturation of the camera at the direct beam location. Spectrometer energy-dispersion range of $64\,$eV was used throughout the camera, with 4x camera binning. We acquired a series of images with slightly different spectrometer energy shifts, with the intensity of the zero-loss peak (ZLP) below the saturation level of the camera. By aligning these images to the ZLP position and then averaging, the weak signal visibility is increased by alignment, while the camera's fibre-coupling background pattern is blurred.

The data were collected using a Hitachi HF-3300 transmission electron microscope equipped with a cold field emission gun \cite{Kaede-san2010} and a CEFID spectrometer developed by CEOS GmbH \cite{CEFID_BookCh_2019} with a TVIPS XF416 CMOS camera. Typically, approximately 1000 spectra were collected with $120\,$ms per spectrum acquisition time and aligned using cross-correlation within CEOS Panta Rhei software. 

The microscope was operated at 300 keV incident electron energy, with an objective lens turned ON and a short camera length $L\approx10\,$cm to include high scattering angles, for example, 1$^{st}$ order Bragg reflections within the spectrometer entrance aperture. Care was taken to avoid sample contamination. The microscope is baked weekly at $\approx$100$^{\circ}$~C, and the column liner tube is regularly cleaned by low-energy electron bombardment \cite{2017_Simon_Contami_Micron,ContamiCharles_2012,TEC_Emi_2020,TEC_Misa_2018,ZONE_sparkle_DavidHoyle2021,UVclean_2011}, keeping the vacuum near the sample $\leq\,5\times 10^{-8}$ torr. Furthermore, the samples were kept at $\geq\,350\,^{\circ}$ C for at least overnight before data collection. Although reducing surface contamination for volume plasmon measurement may not seem as critical as for surface plasmon measurements \cite{2017_Prashant_qEELS_ACSPhotonics}, avoiding (carbonaceous) contamination prevents an undue increase in sample thickness and beam broadening within the sample and multiple inelastic scattering \cite{EELS3, Misa_ThickSample2024,Misa_ThicSample_MSA2022}. 

The free lens control was used to set an initial probe size of $1\,\mu$m on the sample and orient the diffraction pattern with respect to the $q$-EELS slit. The camera length was adjusted by changing the $Z$-position of the sample \cite{midgley1999simple}, which maintains the azimuthal orientation of the diffraction pattern. The angular dispersion of the $q$-EELS spectra was calibrated based on the Bragg spots. The energy dispersion was calibrated using typical procedures for the spectrometer, utilizing signals of known energy loss relative to the ZLP.
\section*{Acknowledgments}
This work was supported by the Office of Naval Research (ONR) under the award number N00014231270. JM acknowledges the NRF \textit{Sejong} Science fellowship funded by the MSIT of the Korean government (RS-2023-00252778). In Canada, we acknowledge the support of the Natural Sciences and Engineering Research Council of Canada (NSERC), RGPIN-2016-04680 and RGPIN-2021-02539. Outstanding support of Hitachi High-Tech, both in Canada and Japan, and CEOS GmbH has made this work possible.
\bibliography{references}
\end{document}